\documentclass[notitlepage,a4paper,aps,prd,onecolumn,superscriptaddress,nofootinbib,groupedaddress]{revtex4}
\usepackage{enumerate}
\usepackage{amsmath}
\usepackage{amsfonts}
\usepackage{amssymb}
\usepackage[utf8]{inputenc}
\usepackage[T1]{fontenc}
\usepackage{mathtools}

\DeclareMathAlphabet{\mathds}{U}{BOONDOX-ds}{m}{n}
\usepackage[dvipsnames]{xcolor}

\usepackage[colorlinks=true]{hyperref}
\usepackage{amsthm}

\theoremstyle{definition}

\theoremstyle{plain}

\newcommand{\dd}{\mathrm{d}}

\newcommand{\lc}[1]{\mathring{#1}}%Levi-Civita connection

\begin{document}
\title{A quick guide to spacetime symmetry and symmetric solutions in teleparallel gravity}

\author{Christian Pfeifer}
\email{christian.pfeifer@zarm.uni-bremen.ee}
\affiliation{ZARM, University of Bremen, 28359 Bremen, Germany}

\begin{abstract}
The notion of spacetime symmetry is essential to describe gravitating physical systems like planets, stars, black holes, or the universe as a whole, since they possess, at least to good approximation, spherical, axial, or spatially homogeneous and isotropic symmetry, respectively. This article gives a quick overview over the known facts on spacetime symmetries in teleparallel gravity. The most general spherical, axial, or spatially homogeneous and isotropic tetrads in Weitzenb\"ck gauge are presented and a brief discussion about symmetric solutions of the anti-symmetric field equations in $f(T,B,\phi,X)$-gravity is given. The article summarizes the authors presentation on this topic at the X. Roberto A. Salmeron School of Physics (EFRAS) in Brasilia, which can be watched online \cite{TalkYT}.
\end{abstract}

\maketitle

%%%%%%%%%%%%%%%%%%%%%%%%%%%%%%%%%%%%%%%%%%%%%%%%%%%%%%%
\section{Teleparallel gravity}\label{sec:telegrav}
Teleparallel gravity offers a fascinating way of reformulating general relativity as a gauge theory of translations, as well as multiple ways to search for generalizations and extensions of Einstein's theory of gravity. These constructions become possible by the use of teleparallel geometry to formulate the dynamics of the gravitational interaction, instead of pseudo-Riemannian geometry.

Teleparallel geometry describes the geometry of a manifold in terms of a tetrad and an independent flat and metric compatible connection; instead of employing a metric and its Levi-Civita connection as it is usually done when one studies general relativity. We will start this overview article in this section by introducing the basic mathematical notions of teleparallel geometry and $f(T,B,\phi,X)$-gravity.

A key ingredient to simplify the derivation of the gravitational field of gravitating physical system is to use their symmetries and to search for symmetric solutions to a theory of gravity. For general relativity, and all theories of gravity which are based on a Lorentzian metric alone, symmetries are implemented by the existence of Killing vector fields, i.e.\ vector fields along which the psuedo-Riemannian geometry is invariant.  In teleparallel gravity, which is based on teleparallel geometry a generalized Killing equation is needed which is a necessary and sufficient condition for the invariance of the geometry under the flow of a set of symmetry generating vector fields. In section \ref{sec:telesymmetry} we will recall the Killing equation for teleparallel geometry and solve it for axial, spherical and homogeneous and isotropic symmetry in the sections \ref{sec:axial}, \ref{sec:spherical} and \ref{sec:cosmo}. 

The solutions to the teleparallel Killing equation then serve as ansatz to solve the field equations of teleparallel theories of gravity, where we will focus on $f(T,B,\phi,X)$-gravity in this article.

Main overview references to teleparallel geometry and teleparallel gravity are: the book \cite{AP} and the reviews \cite{Krssak:2018ywd,Bahamonde:2021gfp}.

%------------------------------------------------------------------%
\subsection{Teleparallel geometry}\label{sec:TPGeom}
Throughout this work, we consider a $4$-dimensional manifold $M$ as spacetime manifold. The fundamental ingredients to a teleparallel geometry of spacetime are:
\begin{itemize}
	\item The tetrads $\{\theta^a\}_{a=0}^3$, i.e.\, $1$-form fields which form a basis of the cotangent spaces at each $p\in M$. They can be expressed in local coordinates $(x)$ and define the spacetime metric $g$
	\begin{align}\label{eq:tetradmetric}
		\theta^a = \theta^a{}_\mu \dd x^\mu,\quad g = g_{\mu\nu} \dd x^\mu \dd x^\nu = \eta_{ab}\theta^a{}_\mu \theta^b{}_\nu\dd x^\mu \dd x^\nu = \eta_{ab} \theta^a \theta^b\,,
	\end{align}
	where $\eta_{ab} = \textrm{diag}(1,-1,-1,-1)$ is the Minkowski metric. We denote the dual vector fields to the tetrads by $\{e_a \}_{a=0}^3$, which can be expressed in local coordinates as $e_a = e_a{}^\mu \partial_\mu$. Their components satisfy $e_a{}^\mu \theta^a{}_\nu = \delta^\mu_\nu$ and $e_a{}^\mu \theta^b{}_\mu = \delta^b_a$.
	
	\item The independent flat and metric compatible connection $1$-form $\omega^a{}_b = \omega^a{}_{b\mu}\dd x^\mu$. It can equally be expressed in terms  of affine connection coefficients as
	\begin{align}\label{eq:affineconnection}
		\Gamma^{\mu}{}_{\nu\rho}(\theta, \omega) = e_a{}^{\mu}(\partial_{\rho}\theta^a{}_{\nu} + \omega^a{}_{b\rho}\theta^b{}_{\nu})\,.
	\end{align}
	The flatness and metric compatibility condition imply that the connection coefficients $ \omega^a{}_{b\mu}$ are given by
	\begin{align}\label{eq:LamdaDLambda}
		 \omega^a{}_{b\mu} = \Lambda^a{}_c\partial_\mu (\Lambda^{-1})^c{}_b\,,
	\end{align}
	where $\Lambda^a{}_b$ satisfy $\eta_{ab}\Lambda^a{}_c \Lambda^b{}_d = \eta_{cd}$, i.e.\ $\Lambda^a{}_b$ is an element of the Lorentz group. Since this teleparallel connection is flat and metric compatible it possesses only one non-vanishing characteristic tensor: its torsion. It serves as building block to define teleparallel theories of gravity from an action principle,
	\begin{align}\label{eq:torsion}
		T^\sigma{}_{\mu\nu}(\theta, \omega) = \Gamma^{\sigma}{}_{\nu\mu} - \Gamma^{\sigma}{}_{\mu\nu}\,.
	\end{align}
\end{itemize}
Thus, teleparallel geometries are manifolds $M$ equipped with a tuple $(\theta^a, \omega^a{}_b)$ that is composed of a tetrad and a teleparallel connection, or, since the connection is locally defined in terms of an element $\Lambda$ of the Lorentz group $SO(1,3)$, by a tuple $(\theta^a, \Lambda^a{}_b)$.

Having fixed a tuple $(\theta^a, \omega^a{}_b)$, or $(\theta^a, \Lambda^a{}_b)$, we can apply a Lorentz transformation $\hat \Lambda$ to obtain a transformed tetrad and Lorentz transformation
\begin{align}
	(\theta^a, \Lambda^a{}_b) \mapsto (\tilde \theta^a, \tilde \Lambda^a{}_b)\,,
\end{align} 
by setting
\begin{align}
	\tilde \theta^a = \theta^b (\hat \Lambda^{-1})^a{}_b,\quad \tilde \Lambda^a{}_b = \hat \Lambda^a{}_c \Lambda^c{}_b\,.
\end{align}
One finds that the geometric objects introduced so far are invariant under the transformation $(\theta^a, \omega^a{}_b)\overset{\hat \Lambda}{\to} (\tilde \theta^a, \tilde \omega^a{}_b)$ by the Lorentz transformation $\hat \Lambda$
\begin{align}
	g_{\mu\nu}(\tilde \theta) = g_{\mu\nu}(\theta),\quad \Gamma^\rho{}_{\mu\nu}(\tilde \theta^a, \tilde \Lambda^a{}_b) ,\quad T^\sigma{}_{\mu\nu}(\tilde \theta,\tilde \omega) = T^\sigma{}_{\mu\nu}(\theta, \omega)\,.
\end{align}
In this sense, local Lorentz transformations of the tuple $(\theta^a, \omega^a{}_b)$ are gauge transformation of the geometry. Thus, every teleparallel theory of gravity that is built from these geometric building blocks possesses this local Lorentz invariance.

A special choice of gauge is the so called the Weitzenb\"ock gauge. By choosing $\hat \Lambda^a{}_b = (\Lambda^{-1})^a{}_b$, any tuple $(\theta^a, \omega^a{}_b)$ can be transformed to the tuple $(\tilde \theta^a, 0)$, since the transformed, teleparallel connection generating Lorentz transformation $\tilde \Lambda$ becomes the constant identity matrix. The Weitzenb\"ock gauge is particularly useful for calculations since the affine connection simplifies to
\begin{align}
	\Gamma^{\mu}{}_{\nu\rho}(\tilde \theta, 0) = \tilde e_a{}^{\mu}\partial_{\rho}\tilde \theta^a{}_{\nu}\,.
\end{align}
 
In the following we will mainly work in Weitzenb\"ock gauge. Before we discuss symmetries of teleparallel geometries, we briefly discuss the field equations of $f(T,B,\phi,X)$-gravity.

%------------------------------------------------------------------%
\subsection{Teleparallel theories of gravity}
A teleparallel theory of gravity is defined from an action for the fundamental gravitational fields $(\theta^a, \omega^a{}_b)$ and additional matter fields $\psi$. It can be expressed as functional on spacetime in the following form
\begin{align}
	S[\theta^a, \omega^a{}_b,\psi] =  S_{TP}[\theta^a, \omega^a{}_b]  + S_M[\theta^a, \omega^a{}_b,\psi] = \frac{1}{2\kappa^2}\int_M \dd^4x |\theta| L(\theta^a, \omega^a{}_b,\partial \theta^a, \partial \omega^a{}_b,....) + S_M[\theta^a, \omega^a{}_b,\psi] \,,
\end{align}
where $|\theta|$ denotes the norm of the determinant of the tetrad, and $L$ is a scalar Lagrange  function, which may depend on the components of the tetrad and the connection, as well as on their derivatives. The matter action is denoted by $S_M[\theta^a, \omega^a{}_b,\psi]$ and can in principle also dependent on all variables. Usually the teleparallel connection only appears, if at all, in the matter action for spinor fields. We will not enter the discussion on mater coupling in teleparallel gravity here, a detailed discussion on thus topic can for example be found here \cite{BeltranJimenez:2020sih}.

Variation of the action with respect to the tetrad yields field equations of the form
\begin{align}
	E_a{}^\mu := (\delta_\theta S_{TP})_a{}^\mu = 2\kappa^2 (\delta_\theta S_{M})_a{}^\mu = 2\kappa^2 \mathcal{T}_a{}^{\mu}\,,
\end{align}
where $\mathcal{T}_a{}^{\mu}$ is the energy momentum tensor generated by the matter fields. It is convenient to multiply this equation by $\theta^a{}_\rho$ and $g_{\mu\sigma}$ to obtain field equations of the form
\begin{align}
	E_{\rho\sigma} = 2 \kappa^2  \mathcal{T}_{\rho\sigma}\,,
\end{align}
which can be decomposed into symmetric and anti-symmetric part 
\begin{align}\label{eq:feq}
	E_{(\rho\sigma)} =  2 \kappa^2 \mathcal{T}_{(\rho\sigma)},\quad E_{[\rho\sigma]} =  2 \kappa^2 \mathcal{T}_{[\rho\sigma]}\,.
\end{align}
Variation with respect to the spin connection components yields equations which are equivalent to the anti-symmetric field equations displayed above \cite{Hohmann:2017duq}. The anti-symmetric part of the energy-momentum tensor is only non-vanishing if matter fields are coupled to the teleparallel connection.

We will focus on $L = f(T,B,\phi,X)$ theories here as an example for a teleparallel theory of gravity for which one can solve the field equations (at least the anti-symmetric part) in the presence of spacetime symmetries. These theories are defined by an arbitrary function of the so-called torsion scalar 
\begin{align}
T=\frac{1}{4}T^{\mu\nu\rho}T_{\mu\nu\rho} + \frac{1}{2}T^{\mu\nu\rho}T_{\rho\nu\mu} - T_{\rho}T^{\rho}\,,\label{defT}
\end{align} 
the boundary term 
\begin{align}
B = \frac{2}{\theta}\partial_{\mu}\left(\theta T^{\sigma}{}_{\sigma}{}^{\mu}\right)\,,
\end{align}
a scalar field $\phi$ and its kinetic energy term $X = \tfrac{1}{2}\partial_\mu\phi\partial^\mu\phi$.  The field equations in Weitzenb\"ock gauge are, see for example~\cite{Bahamonde:2020snl},
\begin{align}\label{fieldeq}
&2\delta_{\nu}^{\lambda}\lc{\Box} f_{B}-2\lc{\nabla}^{\lambda}\lc{\nabla}_{\nu}f_{B} + B f_{B}\delta_{\nu}^{\lambda} + 4\Big[(\partial_{\mu}f_{B})+(\partial_{\mu}f_{T})\Big]S_{\nu}{}^{\mu\lambda}\\
&+4h^{-1}h^{A}{}_{\nu}\partial_{\mu}(h S_{A}{}^{\mu\lambda})f_{T} -  4 f_{T}T^{\sigma}{}_{\mu \nu}S_{\sigma}{}^{\lambda\mu} - f \delta_{\nu}^{\lambda} + \epsilon\, f_{X}\partial^{\lambda}\phi \partial_{\nu}\phi =2\kappa^2 \mathcal{T}_{\nu}^{\lambda}\,,
\end{align}
where a $\mathring{}$ over an object means that it is defined in terms of the Levi-Civita connection of the metric \eqref{eq:tetradmetric}.

This class of  teleparallel theories of gravity contains the ones most studied in the literature:
\begin{itemize}
	\item the teleparallel equivalent of general relativity (TEGR). This teleparallel theory of gravity is dynamically equivalent to general relativity. This means that all tetrads $\theta^a$ which solve the field equations of this theory define metrics $g$ via \eqref{eq:tetradmetric}, that solve the Einstein equations. The reason for this equivalence is that the teleparallel action $L = T$ differs from the Einstein-Hilbert action of general relativity only by the boundary term $B$. In this case the anti-symmetric field equations of \eqref{fieldeq} vanish identically.
	\item $f(T,B)$ gravity and $f(T)$ gravity, for which the field equations \eqref{fieldeq} simplify only marginally. The anti-symmetric field equations for these classes of theories are in general non-vanishing and given by
	\begin{align}
		  E_{[\mu\nu]}:= 4\Big[(\partial_{\rho}f_{B})+(\partial_{\rho}f_{T})\Big]S_{[\mu}{}^{\rho}{}_{\nu]}=\frac{3}{2}T^{\rho}{}_{[\mu\nu}\partial_{\rho]}(f_T+f_B)\,,\label{antitotal}
	\end{align}
	where the subscripts $T$ and $B$ denote the derivative of $f$ with respect to these quantities.
\end{itemize}

In the next section we will discuss how to implement symmetries of a teleparallel geometry and how these help to solve the (anti-symmetric) field equations of $ f(T,B,\phi,X)$ gravity.

%%%%%%%%%%%%%%%%%%%%%%%%%%%%%%%%%%%%%%%%%%%%%%%%%%%%%%%
\section{Spacetime symmetries - the teleparallel killing equation}\label{sec:telesymmetry}
The discussion of symmetries in teleparallel geometry follows \cite{Hohmann:2019nat}, where all details of the derivations can be found. 

A symmetry of a spacetime $M$ is an action $\Phi:G\times M \to M$ of a Lie group $G$ on $M$. This action is given by diffeomorphisms $\Phi_u := \Phi(u,\cdot) : M \to M$ for each $u\in G$, which leave the geometry of $M$ invariant. To make this statement more precise we need to specify what we mean by ''leaving the geometry of $M$ invariant``.

The geometry of $M$ is determined by the affine connection \eqref{eq:affineconnection}. Moreover, the geometry felt by the matter fields which do not couple to the connection is determined by the metric \eqref{eq:tetradmetric} generated by the tetrads. Hence we say that the geometry of $M$ is invariant under the action of a Lie group $G$ if and only if the metric and the affine connection are invariant under all diffeomorphisms $\Phi_u$. Technically this means that their pullback must satisfy
\begin{align}
	(\Phi_u^*g)_{\mu\nu} = g_{\mu\nu} ,\quad (\Phi_u^*\Gamma)^{\rho}{}_{\mu\nu} = \Gamma^{\rho}{}_{\mu\nu}\,.
\end{align}
Infinitesimally, the  diffeomorphisms $\Phi_u$ are generated by vector fields  on $M$ labeled by elements $\xi$ of the Lie algebra $\mathfrak{g}$ of $G$, which leads to the Killing equations for the metric and the connection. A spacetime geometry possesses the symmetry generated by $\mathfrak{g}$ if and only if the Lie derivatives of the metric and the connection with respect to the vector fields $X_\xi$ which generate the diffeomorphisms $\Phi_u$ vanish
\begin{align}
	(\mathcal{L}_{X_\xi}g)_{\mu\nu} &= \lc{\nabla}_\mu X_{\xi\nu}+\lc{\nabla}_\nu X_{\xi\mu}=0\label{eq:killing1a}\\
	(\mathcal{L}_{X_\xi}\Gamma)^{\mu}{}_{\nu\rho} &= \nabla_{\rho}\nabla_{\nu}X^{\mu}_\xi - \nabla_{\rho}(X^{\sigma}_\xi T^{\mu}{}_{\nu\sigma})=0\,,\label{eq:killing1b}
\end{align}
where $\nabla$ is the covariant derivative of the teleparallel connection and $\lc{\nabla}$ the Levi-Civita covariant derivative. 

Using the relation between the teleparallel variables, the tetrad $\theta^a$ and the connection coefficients $\omega^a{}_b$, and the metric and the affine connection, the conditions \eqref{eq:killing1a} and \eqref{eq:killing1b} can be translated into the teleparallel Killing equations, which we display in Weitzenb\"ock gauge
\begin{align}\label{eq:killingTP}
	(\mathcal{L}_{X}\theta)^a{}_{\mu} &= X_{\xi}^{\nu}\partial_{\nu}\theta^i{}_{\mu} + \partial_{\mu}X_{\xi}^{\nu}\theta^i{}_{\nu} =  -\lambda_\xi^a{}_b\theta^b{}_{\mu}\\
	(\mathcal{L}_{X}\omega)^a{}_{b\mu} &= \partial_{\mu}\lambda_\xi^a{}_b =0\,.
\end{align}
In other words, a teleparallel geometry in Weitzenb\"ock gauge is invariant under a Lie group action $G$, if and only if there exists a Lie algebra homomorphism between the symmetry algebra $\mathfrak{g}$ and the Lorentz algebra $\mathfrak{so}(1,3)$, i.e.\ a map $\boldsymbol{\lambda}: \mathfrak{g} \to \mathfrak{so}(1,3)$, such that the tetrad and and the Lie algebra elements $\boldsymbol{\lambda}(\xi)^a{}_b = \lambda_\xi^a{}_b\in \mathfrak{so}(1,3)$ satisfy \eqref{eq:killingTP}.

Let us now derive solutions to the teleparallel Killing equations for certain choices of symmetry groups $G$ and demonstrate how these help us to find solutions for teleparallel theories of gravity.

%%%%%%%%%%%%%%%%%%%%%%%%%%%%%%%%%%%%%%%%%%%%%%%%%%%%%%%
\section{Axial symmetry}\label{sec:axial}
Axial symmetry is encoded in the existence of one Killing vector field which generates the rotations around one axis. This generator forms the Lie algebra $\mathfrak{so}(2)$ and generates the group $SO(2)$. In coordinates $(t,r,\vartheta,\varphi)$ the Killing vector field is simply given by $X_z = \partial_\phi$. 

The first step to solve the teleparallel Killing equations is to determine the Lie algebra homomorphism $\boldsymbol{\lambda}$ that maps the generator $X_z$ into the Lorentz algebra. From the second equation in \eqref{eq:killingTP} we find that it must be constant. There exist two canonical choices for this ismorphism
\begin{align}
	\boldsymbol{\lambda}_1(X_z) = 
	\begin{pmatrix}
	0 && 0 & 0 & 0\\
	0 && 0 & -1 & 0\\
	0 && 1 & 0 & 0\\
	0 && 0 & 0 & 0
	\end{pmatrix}\in \mathfrak{so}(1,3)
	\textrm{ and }
	\boldsymbol{\lambda}_2(X_z) =
	\begin{pmatrix}
	1 && 0 & 0 & 0\\
	0 && 1 & 0 & 0\\
	0 && 0 & 1 & 0\\
	0 && 0 & 0 & 1
	\end{pmatrix}\in \mathfrak{so}(1,3)\,.
\end{align}
Solving the first equation in \eqref{eq:killingTP} for each of them results in the two possibilities for axially symmetric tetrads
\begin{align}
\theta_I^a{}_\mu(t,r,\vartheta,\varphi) 
&= \begin{pmatrix}
C^0{}_0 && C^0{}_1 & C^0{}_2 & C^0{}_3\\
C^1{}_0 \cos\varphi - C^2{}_0 \sin\varphi && C^1{}_1 \cos\varphi - C^2{}_1 \sin\varphi & C^1{}_2 \cos\varphi - C^2{}_2 \sin\varphi & C^1{}_3 \cos\varphi - C^2{}_3 \sin\varphi\\
C^1{}_0 \sin\varphi + C^2{}_0 \cos\varphi && C^1{}_1 \sin\varphi + C^2{}_1 \cos\varphi & C^1{}_2 \sin\varphi + C^2{}_2 \cos\varphi & C^1{}_3 \sin\varphi+ C^2{}_3 \cos\varphi\\
C^3{}_0 && C^3{}_1 & C^3{}_2 & C^3{}_3
\end{pmatrix}\,,\label{eq:axtet1}\\
\theta_{II}^a{}_\mu(t,r,\vartheta)  
&= \begin{pmatrix}
C^0{}_0 && C^0{}_1 & C^0{}_2 & C^0{}_3\\
C^1{}_0 && C^1{}_1 & C^1{}_2 & C^1{}_3\\
C^2{}_0 && C^2{}_1 & C^2{}_2 & C^2{}_3\\
C^3{}_0 && C^3{}_1 & C^3{}_2 & C^3{}_3
\end{pmatrix}\,,\label{eq:axtet2}
\end{align}
where $C^a{}_\mu = C^a{}_\mu(t,r,\vartheta)$. One of the tetrads has a very specific dependence on the angle $\varphi$ while the other is independent of this coordinate. Hence we found two tuples of tetrad and spin connection which are axially symmetric $(\theta_I^a, 0)$ and $(\theta_{II}^a,0)$. 

Observe that the tetrads  $\theta_I^a$ and $\theta_{II}^a$ can be transformed into each other with help of a Lorentz transformation, say $\hat \Lambda$. However, performing such a transformation on  $(\theta_I^a, 0)$ leads to the equivalent tuple $(\theta_{II}^a,\omega^a{}_b(\hat{\Lambda}))$ which is different from $(\theta_{II}^a,0)$. Similarly, when one applies the Lorentz transformation  $\hat \Lambda^{-1}$ to the tuple $(\theta_{II}^a,0)$ one obtains the equivalent tuple $(\theta_I^a, {\omega}^a{}_b(\hat \Lambda^{-1}))$ which is not equivalent to $(\theta_{I}^a,0)$. Thus, non-Weitzenb\"ock gauge axially symmetric teleparallel geometries, which are equivalent to either $(\theta_I^a, 0)$ or $(\theta_{II}^a,0)$ can be obtained by applying a local Lorentz transformations according to the rules discussed in section \ref{sec:TPGeom}.  Further details on the derivation of these tetrads can be found in \cite{Hohmann:2019nat}.

The tetrads \eqref{eq:axtet1} and \eqref{eq:axtet2} lead in general to a an axially symmetric metric with all components. We can use the freedom of redefining the coordinates and fix some of the free functions $C^a{}_\mu$ to find two classes of tetrads 
\begin{align}
\theta_I^a{}_\mu(t,r,\vartheta,\varphi) &= 
\begin{pmatrix}
C^0{}_0 && 0 & 0 & 0\\
0 && C^1{}_1 \cos\varphi & C^1{}_2 \cos\varphi  &  - C^2{}_3 \sin\varphi\\
0 && C^1{}_1 \sin\varphi  & C^1{}_2 \sin\varphi  &  C^2{}_3 \cos\varphi\\
0 && C^1{}_1 C^1{}_2 (C^3{}_2)^{-1} & C^3{}_2 & 0
\end{pmatrix}\,, \label{eq:axtet3}\\
\theta_{II}^a{}_\mu(t,r,\vartheta) &= 
\begin{pmatrix}
C^0{}_0 && 0 & 0 & C^0{}_3\\
0 && C^1{}_1 & C^1{}_2 & 0\\
C^2{}_0 && 0 & 0 & C^2{}_3\\
0 && C^1{}_1 C^1{}_2 (C^3{}_2)^{-1} & C^3{}_2 & 0
\end{pmatrix}\,,\label{eq:axtet4}
\end{align}
which lead to an axially symmetric metric with only one off diagonal component
\begin{align}
	g = g_{tt} dt^2 + g_{rr}dr^2 + g_{\vartheta\vartheta}d\vartheta^s + g_{\varphi\varphi}d\varphi^2 + g_{t\varphi}dtd\varphi\,.
\end{align}
The first branch \eqref{eq:axtet3} is called the regular branch since it has a proper limit to spherically symmetric tetrads, while the second one has not \eqref{eq:axtet4}. The tetrads \eqref{eq:axtet3} and \eqref{eq:axtet4} turn out to be particularly good starting points to find solutions for  $ f(T,B,\phi,X)$ -gravity, see~\cite{Bahamonde:2020snl}, where also all details on the following statements about the solutions can be found.

Assuming no time dependence of the tetrad components ($\partial_t C^a{}_\mu = 0$), for the regular branch \eqref{eq:axtet3} the anti-symmetric field equations \eqref{antitotal} become
\begin{align}
	(\partial_\vartheta f_{T} +\partial_\vartheta f_{B})Q_\vartheta + (\partial_rf_{T} + \partial_rf_{B})Q_r = 0\,,
\end{align}
where
\begin{align}
	Q_\vartheta = \frac{\partial_r C^0{}_0}{ C^0{}_0} - \frac{C^1{}_1}{C^2{}_3} + \frac{\partial_r C^2{}_3}{ C^2{}_3},\quad Q_r = \frac{C^1{}_2 - \partial_\vartheta C^2{}_3}{C^2{}_3} - \frac{\partial_\vartheta C^0{}_0}{ C^0{}_0}\,.
\end{align}
Now there are several options to find solutions to these equations:
\begin{itemize}
	\item Universal solutions for any $f$ by demanding that $Q_\vartheta$ and $Q_r$ vanish. These gives a relation between the tetrad coefficients 
	\begin{align}
	C^1{}_1 = \tfrac{C^2{}_3\partial_rC^0{}_0}{C^0{}_0}+ \partial_rC^2{}_3,\quad C^1{}_2 =\tfrac{C^2{}_3\partial_\vartheta C^0{}_0}{C^0{}_0} +  \partial_\vartheta C^2{}_3\,.
	\end{align}
	This such obtained tetrad lead to a class of metrics which do not contain the famous axially symmetric solutions of general relativity, the Kerr, C or Taub-NUT metric or perturbations of them.
	\item The equations $Q_r=0=(f_{T,\vartheta}+f_{B,\vartheta})$ are solved by a tetrad which generates metrics of the Taub-NUT type
	\begin{align}
	\theta_I^a{}_\mu=\left(
	\begin{array}{cccc}
	\sqrt{\mathcal{A}(r)} & 0 & 0 & \sqrt{\mathcal{A}(r)} \left(C_2+C_1\cos \vartheta\right) \\
	0 & \sqrt{\mathcal{B}(r)} \sin \vartheta \cos \varphi & \sqrt{\mathcal{C}(r)}\cos \vartheta  \cos \varphi & - \sqrt{\mathcal{C}(r)}\sin \vartheta\sin \varphi  \\
	0 & \sqrt{\mathcal{B}(r)} \sin \vartheta \sin \varphi & \sqrt{\mathcal{C}(r)}\cos \vartheta \sin\varphi & \sqrt{\mathcal{C}(r)}\sin \vartheta \cos \varphi  \\
	0 & \sqrt{\mathcal{B}(r)} \cos \vartheta & -\sqrt{\mathcal{C}(r)}\sin \vartheta  & 0 \\
	\end{array}\right)\,.
	\end{align}
	This tetrad is a suitable ansatz for the symmetric field equations to search for teleparallel corrections to the Taub-NUT metric in $ f(T,B,\phi,X)$ -gravity. It is close to the standard spherically symmetric tetrad, as we will see in section \ref{sec:spherical}, and basically only contains one additional term in $t\vartheta$-component.
	\item The solutions to the equations $Q_\vartheta=0=(f_{T,r}+f_{B,r})$ yield metrics which could not be connected to any known solutions of general relativity and whose physical interpretation is unclear so far, which is why we do not discuss this branch here further.
	\item The general case, where none of the terms in the anti-symmetric field equations vanish separately is the most involved one. In order to be able to obtain solutions we make the ansatz that most of the tetrad components are fixed to the values of a tetrad of the Kerr metric
	\begin{align}
		C^0{}_{0}&=\sqrt{1-\frac{2Mr}{\Sigma}}\,,\ \ C^1{}_{1}=\frac{C^3{}_{2}}{\sqrt{\Delta}}\,, \ \ C^1{}_{2}=\sqrt{\Sigma-(C^3{}_{2})^2}\,,\label{kerrA}\\
		C^2{}_{3}&=\sqrt{\sin^2\vartheta\left(\frac{2a^{2}Mr\sin^{2}\vartheta}{\Sigma}+a^2+r^2\right)+(C^0{}_{3})^2}\,,\quad C^0{}_{3}=-\frac{2aMr\sin ^2\vartheta}{\sqrt{\Sigma(\Sigma-2Mr)}} \,,\label{kerrB}
	\end{align}
	where $a$ is the angular momentum parameter, $\Sigma = r^2+a^2\cos^2\vartheta$ and  $\Delta=r^2-2Mr+a^{2}$. By making a power series ansatz for the remaining tetrad component 
	\begin{align}
		C^3{}_2 = r \sin\vartheta + \mathcal{A}(r,\vartheta) a^2 + \mathcal{O}(a^3)
	\end{align}
	one can solve the anti-symmetric field equations to second order in $a$ with
	\begin{align}
		\mathcal{A}(r,\vartheta)&=\frac{\sin \vartheta  \cos ^2\vartheta  \left(4 \mu ^5+6 \mu ^2 r^{3/2}+r^{5/2}+4 \mu  r^2+\mu ^4 \sqrt{r}-16 \mu ^3 r\right)}{2 \mu ^2 r^{3/2} \left(-\mu ^2-4 \mu  \sqrt{r}+r\right)}+\mathcal{C}(r)F_1(r,\vartheta)+\mathcal{D}(r)F_2(r,\vartheta)\,,
	\end{align}
	where $\mu = \sqrt{r - 2M}$, and $\mathcal{C}(r)$ and $\mathcal{D}(r)$ are arbitrary functions (related to the integration of the differential equation) and $F_1(r,\vartheta)$ and $F_2(r,\vartheta)$ are specific functions which are related to the Legendre function of the first and second kind, see~\cite{whittaker2020course}. Hence, so far it is only possible to find a good ansatz for a$ f(T,B,\phi,X)$-gravity extension of Kerr geometry perturbatively. The search for a non-pertutbative teleparallel correction ot Kerr geometry is an ongoing research project.
\end{itemize}

The study of solutions of the anti-symmetric field equations for the second branch axial tetrad \eqref{eq:axtet4} is also in its infancy, which is why it will not be discussed further here. The interested reader finds further details in \cite[Sec.~V]{Bahamonde:2020snl}.

%%%%%%%%%%%%%%%%%%%%%%%%%%%%%%%%%%%%%%%%%%%%%%%%%%%%%%%
\section{Spherical symmetry}\label{sec:spherical}
Spherically symmetric geometries are invariant under the group action of $SO(3)$. To implement this invariance we add two further Killing vector fields to the axial symmetric one, so that all together they form the $\mathfrak{so}(3)$ algebra \cite{Hohmann:2019nat}. In coordinates $(t,r,\vartheta,\varphi)$ they take the form
\begin{align}
	X_z = \partial_\varphi,
	\quad 
	X_y = -\cos\varphi\partial_{\vartheta} + \frac{\sin\varphi}{\tan\vartheta}\partial_{\varphi},
	\quad 
	X_x = \sin\varphi\partial_{\vartheta} + \frac{\cos\varphi}{\tan\vartheta}\partial_{\varphi}\,.
\end{align}
As in the previous case we must find a homomorphism, which maps these generators of $\mathfrak{so}(3)$ into $\mathfrak{so}(1,3)$. In this case there exists only one such homomorphism which is given by
\begin{align}
 \boldsymbol{\lambda}(X_z) = 
 \begin{pmatrix}
 &0 & 0 & 0 & 0\\
 &0 & 0 & -1 & 0\\
 &0 & 1 & 0 & 0\\
 &0 & 0 & 0 & 0
 \end{pmatrix}\,,
 \quad
 \boldsymbol{\lambda}(X_y) =
 \begin{pmatrix}
 0 && 0 & 0 & 0\\
 0 && 0 & 0 & 1\\
 0 && 0 & 0 & 0\\
 0 && -1 & 0 & 0
 \end{pmatrix}\,,
 \quad
 \boldsymbol{\lambda}(X_x) =
 \begin{pmatrix}
 0 && 0 & 0 & 0\\
 0 && 0 & 0 & 0\\
 0 && 0 & 0 & -1\\
 0 && 0 & 1 & 0
 \end{pmatrix}\,.
\end{align}
Solving the remaining teleparallel killing equation for the tetrad yields the most general spherically symmetric tetrad
\begin{align}\label{eq:sphtet}
	\theta^a{}_\mu = 
	\begin{pmatrix}
	C_1 & C_2 & 0 & 0 \\
	C_3 \sin\vartheta \cos\varphi & C_4 \sin\vartheta \cos\varphi & C_5 \cos\vartheta \cos \varphi - C_6 \sin\varphi  & -\sin\vartheta (C_5 \sin\varphi + C_6 \cos\vartheta \cos\varphi)  \\
	C_3 \sin\vartheta \sin\varphi & C_4 \sin\vartheta \sin\varphi & C_5 \cos\vartheta \sin \varphi + C_6 \cos\varphi  & \sin\vartheta (C_5 \cos\varphi - C_6 \cos\vartheta \sin\varphi) \\
	C_3 \cos\vartheta & C_4 \cos\vartheta & - C_5 \sin\vartheta & C_6 \sin^2\vartheta\\
	\end{pmatrix}
\end{align}
where $C_i = C_i(t,r)$. The resulting metric is the standard spherically symmetric metric with one off diagonal term that can be set to zero by a redefinition of the t coordinate
\begin{align}\label{eq:metricsph}
g= (C_1^2-C_3^2) dt^2 - (C_4^2-C_2^2) dr^2 - (C_5^2+C_6^2) (d\vartheta^2 + sin^2\vartheta d\varphi^2) - (C_3C_4 - C_1 C_2) dt dr\,.
\end{align}

Employing \eqref{eq:sphtet} in the anti-symmetric field equations of $f(T,B,\phi,X)$ -gravity yields two non-trivial expression
\begin{align}
C_3 C_5 (f'_T+f'_B)=0\,,\quad C_1 C_6(f_T'+f_B')=0\,,
\end{align}
which can be solved by the following choices:
\begin{itemize}
	\item $C_2 = C_6 = 0$ and fixing the $t$ and $r$-coordinate by setting $C_2=0$ and $C_5 = \xi r$, where $\xi=\pm1$. This yields the two real tetrads
	\begin{align}
		\theta^a_{\pm}{}_\mu = 
		\begin{pmatrix}
		C_1 & 0 & 0 & 0 \\
		0 & C_4 \sin\vartheta \cos\varphi & r \xi  \cos\vartheta \cos \varphi   & - r \xi \sin\vartheta sin\varphi \\
		0 & C_4 \sin\vartheta \sin\varphi & r \xi \cos\vartheta \sin \varphi   &  r \xi \sin\vartheta \cos\varphi  \\
		0 & C_4 \cos\vartheta & - r \xi \sin\vartheta & 0
	\end{pmatrix}\,.
	\end{align} 
	\item Setting $C_1 = C_5 = 0$ and fixing the $t$ and $r$-coordinate by setting $C_4=0$ and $C_6 = \chi r$, where $\chi=\pm1$, leads to two complex tetrads
	\begin{align}
		\theta^a_\pm{}_\mu = 
		\begin{pmatrix}
		0 & i C_2 & 0 & 0 \\
		i C_3 \sin\vartheta \cos\varphi & 0 &  - r \chi \sin\varphi  & -r \chi \sin\vartheta  \cos\vartheta \cos\varphi \\
		i C_3 \sin\vartheta \sin\varphi & 0 & r \chi \cos\varphi  & r \chi \sin\vartheta \cos\vartheta \sin\varphi \\
		i C_3 \cos\vartheta & 0 & 0 & r \chi \sin^2\vartheta
	\end{pmatrix}\,.\label{eq:sphcomp}
	\end{align}
	The complexification of the tetrad is necessary to preserve the signature of the metric \eqref{eq:metricsph}. The metric itself is diagonal and real.
\end{itemize}
These four tetrads are the spherically symmetric ones which serve as starting point to search for spherically symmetric solutions of the symmetric field equations of  $f(T,B,\phi,X)$-gravity. It is possible to find numerous perturbative solutions in $f(T)$ and $f(T,B)$ gravity, which can be interpreted as teleparallel corrections to Schwarzschild geometry and can be tested against observations like the shadow of black holes, the perihelion shift of stars orbiting a black hole, lensing effects and the shapiro delay~\cite{DeBenedictis:2016aze,Bahamonde:2019zea,Bahamonde:2020vpb,Pfeifer:2021njm,Bahamonde:2021srr}. In general the complex tetrads and the two different real tetrads lead to very different phenomenology.

For the complex tetrad \eqref{eq:sphcomp} it is even possible to find an exact solution to teleparallel Born-Infeld gravity~\cite{Ferraro:2008ey,Bohmer:2019vff}
\begin{align}
	f(T,B,\phi,X) = f(T) = \alpha \left(\sqrt{1+\tfrac{2T}{\alpha}}-1\right)\,.
\end{align}
It is the first non-perturbative spherically symmetric solution to a teleparallel theory of gravity (beyond Schwarzschild geometry in TEGR), and is given by, see \cite{Bahamonde:2021srr},
\begin{align}
	(C_3)^2 &= \frac{a_1^2 }{r}\Big[\sqrt{\alpha } (a_0 \alpha +r)-2 \tan ^{-1}\left(\frac{\sqrt{\alpha } r}{2}\right)\Big]\\
	(C_2)^2 &= \frac{\alpha ^{5/2} r^5}{(4 + r^2 \alpha)^2}\Big[\sqrt{\alpha } (a_0 \alpha +r)-2 \tan ^{-1}\left(\frac{\sqrt{\alpha } r}{2}\right)\Big]^{-1}\,.
\end{align}
For this solution to give an asymptotically flat metric one needs to fix the constant of integration $a_1 = \sqrt{\alpha^{-1}}$. If one wants moreover, that this solution becomes Schwarzschild geometry for $\alpha\to\infty$, one needs to fix the second constant of integration as $a_0 = -\frac{2M}{\alpha}$.

%%%%%%%%%%%%%%%%%%%%%%%%%%%%%%%%%%%%%%%%%%%%%%%%%%%%%%%
\section{Homogeneous and isotropic symmetry}\label{sec:cosmo}
Last but not least we enlarge the symmetry algebra further to the full cosmological spatial homogeneous and isotropic symmetry, which is generate by the vector fields
\begin{align}
X_z &= \partial_\phi\,,
\quad 
X_y = -\cos\varphi\partial_{\vartheta} + \frac{\sin\varphi}{\tan\vartheta}\partial_{\varphi}\,,
\quad 
X_x = \sin\varphi\partial_{\vartheta} + \frac{\cos\varphi}{\tan\vartheta}\partial_{\varphi}\,,\\
X_1 &= \chi\sin\vartheta\cos\varphi\partial_r + \frac{\chi}{r}\cos\vartheta\cos\varphi\partial_{\vartheta} - \frac{\chi\sin\varphi}{r\sin\vartheta}\partial_{\varphi}\\
X_2 &= \chi\sin\vartheta\sin\varphi\partial_r + \frac{\chi}{r}\cos\vartheta\sin\varphi\partial_{\vartheta} + \frac{\chi\cos\varphi}{r\sin\vartheta}\partial_{\varphi}\\
X_3 &= \chi\cos\vartheta\partial_r - \frac{\chi}{r}\sin\vartheta\partial_{\vartheta}\,,
\end{align}
where $\chi = \sqrt{1-kr^2},\ u = \sqrt{k}$ and $k$ is the spatial curvature parameter. The homorphism $\boldsymbol{\Lambda}$ depends on if $k<0$, $k>0$ or $k=0$. A complete classification of the possible homorphisms can be found in \cite{Hohmann:2020zre}. Solving the teleparallel Killing equation \eqref{eq:killingTP} for the tetrad for the different possibilities leads to two branches of spatially homogeneous and isotropic tetrads in Wetzenb\"ock gauge:
\begin{align}
\theta_I^a{}_{\mu} &=
\begin{pmatrix}
N \chi & i u A r\chi^{-1} & 0 & 0\\
i u N r \sin \vartheta \cos \varphi & A \sin \vartheta \cos \varphi  & A r \cos\vartheta \cos\varphi  & - A r \sin\vartheta \sin\varphi\\
i u N r \sin \vartheta \sin \varphi & A \sin \vartheta \sin \varphi  & A r \cos\vartheta \sin\varphi  &   A r \sin\vartheta \cos\varphi\\
i u N r \cos \vartheta & A \cos \vartheta & - A r \sin\vartheta  & 0
\end{pmatrix}\,,\label{eq:cosmo1}\\
{\theta}_{II}^a{}_{\mu} &=
\begin{pmatrix}
N  & 0 & 0 & 0\\
0 & A \sin \vartheta \cos \varphi\chi^{-1}  & A r (\chi \cos\vartheta \cos\varphi + u r \sin\varphi)  & - A r \sin\vartheta ( \chi \sin\varphi - u r \cos\vartheta \cos\varphi )\\
0 & A \sin \vartheta \sin \varphi \chi^{-1}  & A r (\chi \cos\vartheta \sin\varphi + u r \cos\varphi)  &  A r \sin\vartheta ( \chi \cos\varphi + u r \cos\vartheta \sin\varphi )\\
0 & A \cos \vartheta\chi^{-1} & - A r \chi \sin\vartheta  & - u r^2 \sin^2\vartheta
\end{pmatrix}\,.\label{eq:cosmo2}
\end{align}
Both branches of tetrads can be real or complex depending on the sign of $k$. They coincide for $k=0$.

The most interesting fact about these tetrads is, that they solve the anti-symmetric field equations for any teleparallel theory of gravity. The reason is that for tetrads which obey the symmetry conditions, all derived tensors like the torsion and its covariant derivatives satisfy the symmetry conditions. Hence the anti-symmetric field equation of any teleparallel theory of gravity are given by an anti-symmetric spatially homogeneous and isotropic $(0,2)$-tensor field~$E_{[\mu\nu]}$. However, it can be shown, see \cite{Hohmann:2019nat}, that such a tensor field can only have vanishing components, and hence~$E_{[\mu\nu]}=0$. Thus the tetrads \eqref{eq:cosmo1} and \eqref{eq:cosmo2} are universal solutions of the anti-symmetric field equations in cosmology in any teleparallel theory of gravity.

%%%%%%%%%%%%%%%%%%%%%%%%%%%%%%%%%%%%%%%%%%%%%%%%%%%%%%%
\section{Conclusion}\label{sec:conc}
Spacetime symmetries give important simplifications in the search for solutions of teleparallel theories of gravity. To apply spacetime symmetries in teleparallel gravity we discussed the teleparallel Killing equation \eqref{eq:killingTP}, which we solved for the most important classes of symmetries:
\begin{itemize}
	\item In axial symmetry, there exist two independent classes of Weitzenb\"ock tetrads. One with a specific dependence on the $\varphi$ coordinate \eqref{eq:axtet1} (or \eqref{eq:axtet3} in its minimal version), and one that is independent of the $\varphi$ coordinate \eqref{eq:axtet2} (or \eqref{eq:axtet4} in its minimal version);
	\item in spherical symmetry there is only one class of Weitzenb\"ock tetrads, which is given by \eqref{eq:sphtet};
	\item while there exists again two classes of spatially homogeneous and isotropic tetrads \eqref{eq:cosmo1} and \eqref{eq:cosmo2}, for which it depends on the sign of the spatial curvature parameter $k$ if they are real or complex.
\end{itemize}
These tetrads are good starting points to find solutions to the field equations in teleparallel theories of gravity. 

The homogeneous and isotropic ones are universal solutions to the anti-symmetric field equations for all teleparallel theories of gravity. In spherical symmetry there exist four branches of solutions in $f(T,B,\phi,X)$-gravity, while in axial symmetry numerous solutions are possible, whose physical viability must be investigated case by case.

In ongoing and future research projects the tetrads presented here will help to find physical solutions in various teleparallel theories of gravity, also beyond $f(T,B,\phi,X)$-gravity.

%%%%%%%%%%%%%%%%%%%%%%%%%%%%%%%%%%%%%%%%%%%%%%%%%%%%%%%
\begin{acknowledgments}
C.P. was funded by the Deutsche Forschungsgemeinschaft (DFG, German Research Foundation) - Project Number 420243324. Moreover, I like to thank Prof. Sérgio Costa Ulhoa for the kind invitation to talk about spacetime symmetries in teleparallel gravity at the X. Roberto A. Salmeron School of Physics (EFRAS), and huge thanks to my collaborators on the articles which I summarized here for the fantastic work together.
\end{acknowledgments}

%%%%%%%%%%%%%%%%%%%%%%%%%%%%%%%%%%%%%%%%%%%%%%%%%%%%%%%
\bibliographystyle{utphys}
\bibliography{SymmSummTP}

\end{document}